\documentclass[11pt,a4paper]{article}

\usepackage{latexsym,amssymb, amsthm}
\usepackage[centertags]{amsmath}
\usepackage{epsfig,pifont,cite}
\usepackage[inactive]{Srcltx}  

\usepackage{graphics,helvet,times,mathptm,epic,eepic}

\usepackage{color}

\usepackage{newlfont}

\usepackage{amsfonts,bbm}

\newenvironment{keywords}{ \noindent {\small\bf Key Words}:}{ }

\def\G1{\hbox{$\displaystyle{\mbox{\ding{172}}}$}}

\def\bd{\begin{description}}
\def\ed{\end{description}}

\def\beq{\begin{equation}}
\def\eeq{\end{equation}}
\def\bea{\begin{eqnarray}}
\def\eea{\end{eqnarray}}
\def\beas{\begin{eqnarray*}}
\def\eeas{\end{eqnarray*}}

\newtheorem{theorem}{Theorem}

\theoremstyle{remark}
\newtheorem{example}{Example}

\begin{document}

\title{\textbf{\textsc{Using blinking fractals for mathematical
 modeling  of   processes
 of growth  in  biological systems}}}

\newcommand{\nms}{\normalsize}
\author{\\ {   \bf Yaroslav D. Sergeyev\footnote{Yaroslav D.
Sergeyev, Ph.D., D.Sc., is   Distinguished
 Professor  at the University of Calabria, Rende, Italy.
 He is also Full Professor (part-time contract) at the N.I.~Lobatchevsky State University,
  Nizhni Novgorod, Russia and   Affiliated  Researcher at the Institute of High Performance Computing
   and Networking of the National Research Council of Italy.  His scientific interests include infinity computing,
   fractals, numerical  analysis, parallel computations,
   and number theory.}\,\,  \footnote{The author thanks F.M.H.~Khalaf for his kind help in drawing figures.}}\\[-4pt]
      \nms Dipartimento di Elettronica, Informatica e Sistemistica,\\[-4pt]
       \nms   Universit\`a della Calabria,\\[-4pt]
       \nms 87030 Rende (CS)  -- Italy\\ \\[-4pt]
        \nms tel./fax: +39 0984 494855\\[-4pt]
       \nms http://wwwinfo.deis.unical.it/$\sim$yaro\\[-4pt]
         \nms {\tt  yaro@si.deis.unical.it }
}

\date{}

\maketitle

\begin{abstract}
Many biological processes and objects can be described by
fractals. The paper uses a new type of objects -- blinking
fractals -- that are not covered by traditional theories
considering dynamics of self-similarity processes. It is shown
that both traditional and blinking fractals can be successfully
studied by a recent approach allowing one to work numerically with
infinite and infinitesimal numbers.  It is shown that blinking
fractals can be applied for modeling complex processes of growth
of biological systems including their season changes.  The new
approach allows one to give various quantitative characteristics
of the obtained blinking fractals models of biological systems.
 \end{abstract}


\section{Introduction}
\label{s1}

\begin{figure}[t]
  \begin{center}
      \epsfig{ figure = 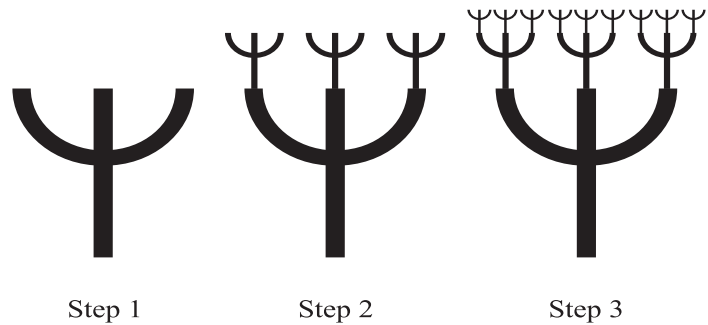, width = 3.6in, height = 1.7in,  silent = yes }
    \caption{A simple fractal model of a tree.}
 \label{Biology_1}
  \end{center}
\end{figure}

Fractals have been very well studied during the last few decades
and have been used  in various scientific fields including biology
to model complex systems   (see numerous applications given in
\cite{Devaney,Falconer,Hastings_Sugihara,fractals,Strongin_Sergeyev}).
The fractal objects are constructed by using the principle of
self-similarity: a given basic figure (some times slightly
modified in time) infinitely many times repeats itself in several
copies. A simple example of such a construction is shown in
Fig.~\ref{Biology_1}. The basic figure shown in Step~1 is then
repeated and already at Step~3 can be viewed as a simple model of
a tree.

The introduction of fractals has allowed people to describe
complex systems having a fractal structure in an elegant and very
efficient way, to construct their computational models, and to
study them. However, it is important to mention that mathematical
analysis of fractals (except, of course, a very well developed
theory of fractal dimensions, see
\cite{Devaney,Falconer,Hastings_Sugihara,fractals}) very often
continues to have mainly a qualitative character. For example,
tools for a quantitative analysis of fractals at infinity are not
very rich yet (e.g., even for one of the mostly studied fractals
-- Cantor's set -- we are not able to count the number of
intervals composing the set at infinity).

Nowadays, fractals usually are used to describe   objects (see,
e.g., a tree presented in Fig.~\ref{Biology_1}, Step~3) where
\textit{one} basic figure often called \textit{generator} can be
determined; they are rarely used for modeling processes where
appearance of the studied objects is changed in time without
preserving the generator. For example, the model from
Fig.~\ref{Biology_1} cannot be used to describe a tree if we take
into account season changes because in summer the tree has green
leafs, in autumn the leafs are yellow, in winter there are no
leafs at all and branches of the tree are under the snow. Thus, we
are not able to distinguish one basic figure that maintains its
form during the whole process and can be observed at all four
seasons.

In nature, there exist processes and objects that evidently are
very similar to classical fractals but cannot be covered by the
traditional approaches because several self-similarity mechanisms
participate in the process of their construction. A  new
methodology (see \cite{Sergeyev,Lagrange}) allowing one not only
to study traditional fractals but also to introduce and to
investigate   a new class of objects -- \textit{blinking fractals}
-- that are not covered by traditional theories studying
self-similarity processes can be used for describing such
processes.

The new methodology  allowing one to work with such processes  can
be found in a rather comprehensive form in
\cite{informatica,Lagrange} downloadable from \cite{www} (see also
the monograph \cite{Sergeyev} written in a popular manner and the
survey \cite{Lolli} where the new methodology  is considered in a
historical panorama of views on infinities and infinitesimals).
Numerous examples of the usage of the   methodology
\cite{informatica,Lagrange} for mathematical modelling in several
fields can be found in
\cite{DeLeone,MM_bijection,Margenstern,Rosinger2,Sergeyev,chaos,Dif_Calculus,Menger,Korea,first,
Riemann,Sergeyev_Garro,DeBartolo,Zhigljavsky,Zilinskas}. The goal
of the entire operation is to propose a way of thinking that would
allow us to work with finite, infinite, and infinitesimal numbers
in the same way and to create mathematical models better
describing the natural phenomena.

In this paper,   blinking fractals (see \cite{chaos}) are used to
model season changes and processes of growth in biological
systems. The paper not only proposes  such a modeling but also
describes mathematical tools allowing one to study the properties
of processes of growth in the limit even in the situations where
various kinds of divergency take place.

The rest of the paper is organized as follows. Section~\ref{s2}
introduces the notion of blinking fractals, presents some
examples, and briefly introduces the methodology used in the
further investigation.   In Section~\ref{s4}, it is shown how the
blinking fractals can be investigated by using the infinite  and
infinitesimal numbers from \cite{informatica,Lagrange}.
Section~\ref{s5} shows how processes of growth of biological
systems can be modelled by using the blinking fractals,
particularly, the new applied approach to infinity is used to
study a model of the growth of a forest. Section~\ref{s6}
concludes the paper.

In conclusion of the Introduction I am happy   to dedicate this
paper to Professor Jonas Mockus in the occasion of his 80 year
jubilee.

\section{Blinking fractals and infinite integers}
\label{s2}



\begin{figure}[t]
  \begin{center}
      \epsfig{ figure = 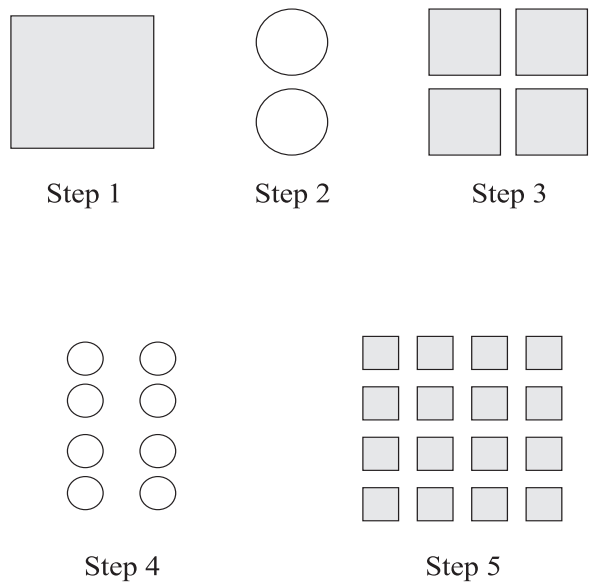, width = 3in, height = 3in,  silent = yes }
    \caption{Results of the first five iterations.}
 \label{Biology_3}
  \end{center}
\end{figure}

Before going to a general definition of blinking fractals let us
consider a process shown in Fig.~\ref{Biology_3}. At the first
moment we see   a grey square with the side equal to~1. At the
second moment we see two  white circles with the diameter equal to
$\frac{1}{2}$. Then  each white circle is substituted by to grey
squares $\frac{1}{2}$ on side.   This process of substitution
continues in time as it shown in Fig.~\ref{Biology_3}.

\begin{figure}[t]
  \begin{center}
      \epsfig{ figure = 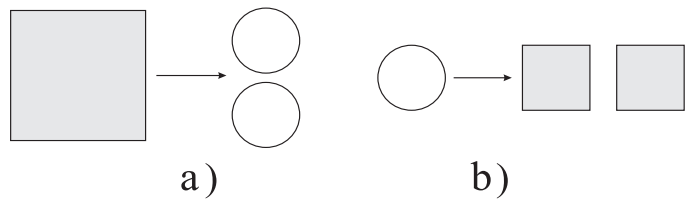, width = 3.5in, height = 1in,  silent = yes }
    \caption{At each even iteration every square with a side equal to $h$ is
     substituted by two circles having the diameter~$\frac{h}{2}$. At each odd iteration every circle
     with a diameter~$d$ is substituted by two squares with $d$ on side.}
 \label{Biology_4}
  \end{center}
\end{figure}

\begin{figure}[t]
  \begin{center}
      \epsfig{ figure = 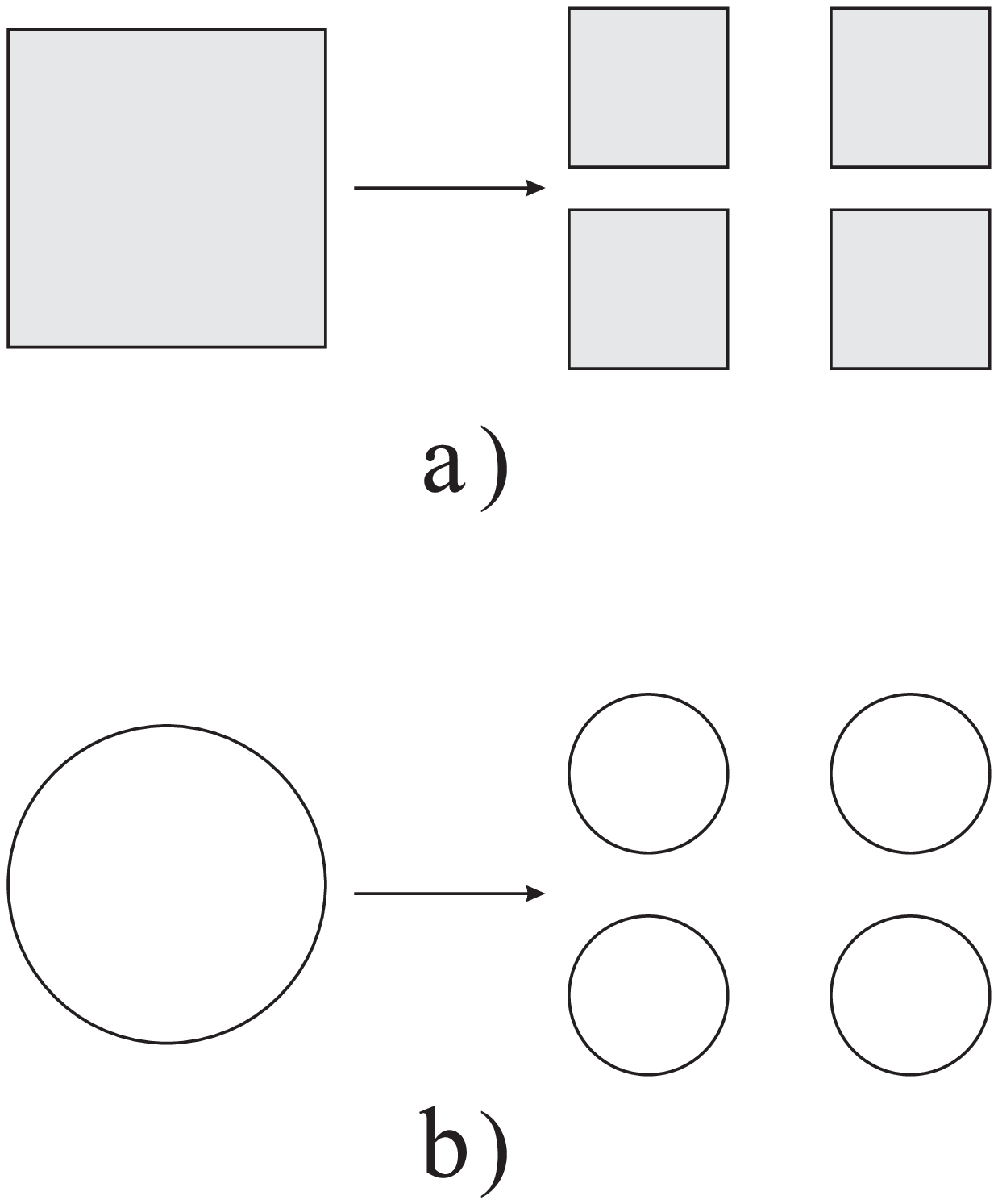, width = 2.3in, height = 2.6in,  silent = yes }
    \caption{Two traditional fractal processes that can be extracted from the blinking fractal process shown in Fig.~\ref{Biology_3}.}
 \label{Biology_5}
  \end{center}
\end{figure}

It is clear that the process shown in Fig.~\ref{Biology_3} is not
a fractal process because  at each even  iteration squares are not
transformed in smaller copies of themselves but in circles (see
Fig.~\ref{Biology_3} and Fig.~\ref{Biology_4}, left). Analogously,
at odd iterations circles are transformed in squares instead of
smaller circles (see Fig.~\ref{Biology_3} and
Fig.~\ref{Biology_4}, right). Thus, the process shown in
Fig.~\ref{Biology_3} is a mixture of two fractal processes with
the rules shown in Fig.~\ref{Biology_5}: the first of them works
with grey squares and the second with white circles.

Traditional approaches are not able to say anything about the
behavior of this process from Fig.~\ref{Biology_3} at infinity.
Does there exist a limit object of this process? If it exists,
what can we say about its structure? Does it consist of white
circles or grey squares and how many of them take part of this
limit object?   All these questions remain without answers if
traditional mathematical tools are used for analysis of such
processes.

In this paper, we give answers to these questions by using a new
approach developed  in \cite{Sergeyev,informatica,Lagrange} for
dealing with infinite, finite, and infinitesimal numbers. The new
methodology will be applied to study traditional fractals and new
objects constructed using the principle of self-similarity with an
infinite cyclic application of \textit{several fractal rules}.
These objects are called hereinafter \textit{blinking fractals}.

Usually, when mathematicians deal with infinite objects (sets or
processes) it is supposed that human beings are able to execute
certain operations infinitely many times (see
\cite{Cantor,Conway,Loeb,Robinson}). For example, in a fixed
numeral system it is possible to write down a numeral\footnote{ We
remind that \textit{numeral}  is a symbol or group of symbols that
represents a \textit{number}. The difference between numerals and
numbers is the same as the difference between words and the things
they refer to. A \textit{number} is a concept that a
\textit{numeral} expresses. The same number can be represented by
different numerals. For example, the symbols `7', `seven', and
`VII' are different numerals, but they all represent the same
number.} with \textit{any} number of digits. However, this
supposition is an abstraction   because we live in a finite world
and all human beings and/or computers finish operations they have
started.

The new computational paradigm introduced in
\cite{Sergeyev,informatica,Lagrange} does not use this abstraction
and, therefore, is closer to the world of practical calculations
than traditional approaches. Its strong computational character is
enforced also by the fact  that the first simulator of the
Infinity Computer able to work with infinite, finite, and
infinitesimal numbers introduced in
\cite{Sergeyev,informatica,Lagrange} has been already realized
(see \cite{www,Sergeyev_patent}).

  The main idea of the new approach consists of the possibility
to measure infinite and infinitesimal quantities by different
(infinite, finite, and infinitesimal) units of measure. A new
infinite unit of measure   has been introduced for this purpose as
the number of elements of the set $\mathbb{N}$ of natural numbers.
The new number is  called \textit{grossone} and is expressed by
the numeral \G1. It is necessary to stress immediately that \G1 is
not related either to non-standard analysis or to Cantor's
$\aleph_0$ and $\omega$. Particularly, \G1 has both cardinal and
ordinal properties as usual finite natural numbers (see
\cite{informatica,first,Lagrange}).  In fact, infinite positive
integers that can be viewed through numerals including grossone
can be interpreted in the terms of the number of elements of
certain infinite sets.

For instance, the set of even numbers has $\frac{\G1}{2}$ elements
and the set of integers has $2\mbox{\G1\small{+}}1$ elements (see
\cite{informatica,first,Lagrange}). Thus, the new numeral system
allows one to distinguish within countable sets many different
sets having the different infinite number  of elements.
Analogously, within uncountable sets it is possible to distinguish
sets having, for instance, $2^{\mbox{\tiny{\G1}}}$ elements,
$10^{\mbox{\tiny{\G1}}}$ elements,   and even
$\G1^{\mbox{\tiny{\G1}}}-1$, $\G1^{\mbox{\tiny{\G1}}}$, and
$\G1^{\mbox{\tiny{\G1}}}+1$ elements and to show  (see
\cite{informatica,first,Lagrange}) that
 \[
\frac{\G1}{2} < \G1 < 2\mbox{\G1\small{+}}1 <
2^{\mbox{\tiny{\G1}}} < 10^{\mbox{\tiny{\G1}}} <
\G1^{\mbox{\tiny{\G1}}}-1 < \G1^{\mbox{\tiny{\G1}}}<
\G1^{\mbox{\tiny{\G1}}}+1.
\]

It is worthwhile to emphasize that since \G1, on the one hand, and
$\aleph_0$ (and $\omega$), on the other hand, belong to different
mathematical languages working with different theoretical
assumptions, they cannot be used together. Analogously, it is not
possible to use together Piraha's `many' (see the primitive
numeral system described in \cite{Gordon}) and the modern numeral
4.

Formally, grossone is introduced as a new number by describing its
properties postulated by the \textit{Infinite Unit Axiom} (IUA)
(see \cite{Sergeyev,informatica}). This axiom is added to axioms
for real numbers similarly to addition of the axiom determining
zero to axioms of natural numbers when integer numbers are
introduced. It is important to emphasize that we speak about
axioms for real numbers in the following applied sense: axioms do
not define real numbers, they just define formal rules of
operations with  numerals in given  numeral systems (tools of the
observation)  reflecting so certain (not all) properties of the
object of the observation, i.e., properties of real numbers.

Inasmuch as it has been postulated that grossone is a number,  all
other axioms for numbers hold for it, too. Particularly,
associative and commutative properties of multiplication and
addition, distributive property of multiplication over addition,
existence of   inverse  elements with respect to addition and
multiplication hold for grossone as for finite numbers. This means
that  the following relations hold for grossone, as for any other
number
 \beq
 0 \cdot \G1 =
\G1 \cdot 0 = 0, \hspace{3mm} \G1-\G1= 0,\hspace{3mm}
\frac{\G1}{\G1}=1, \hspace{3mm} \G1^0=1, \hspace{3mm}
1^{\mbox{\tiny{\G1}}}=1, \hspace{3mm} 0^{\mbox{\tiny{\G1}}}=0.
 \label{3.2.1}
       \eeq
The introduction of the new numeral allows us to use it for
construction of various new numerals expressing infinite and
infinitesimal numbers  and to operate with them as with usual
finite constants. As a consequence, the numeral $\infty$   is
excluded from our new mathematical language (together with
numerals $\aleph_0, \aleph_1, \ldots$  and $\omega$). In fact,
since we are able now to express explicitly different infinite
numbers, records of the type $\sum_{i=1}^{\infty}a_i$   are not
sufficiently precise. It becomes necessary not only to say that
$i$ goes to infinity, it is necessary to indicate to which point
in infinity (e.g., $\G1, 7\G1-1, 3\G1^2+4$, etc.) we want to sum
up. Note that for sums having a finite number of items the
situation is the same: it is not sufficient to say that the number
of items in the sum is finite, it is necessary to indicate
explicitly the number of items in the sum.

The appearance of new numerals expressing infinite and
infinitesimal numbers   gives us a lot of new possibilities. For
example, it becomes possible to develop a Differential Calculus
(see \cite{Dif_Calculus}) for functions that can assume finite,
infinite, and infinitesimal values and can be defined over finite,
infinite, and infinitesimal domains avoiding indeterminate forms
and divergences (all these concepts just do not appear in the new
Calculus). This approach allows us to work with derivatives and
integrals that can assume not only finite but infinite and
infinitesimal values, as well. Infinite and infinitesimal numbers
are not auxiliary entities in the new Calculus, they are full
members in it and can be used in the same way as finite constants.

 These numerals
give us a possibility to study traditional fractals and blinking
fractals at different points of infinity and to use them for
modeling the nature. In this section, we  investigate the blinking
fractal introduced in Section~\ref{s2} and infinite sequences
will be used for this goal. Naturally, we need first to understand
what can we say about the infinite sequences using the new
mathematical language.

\section{Quantitative analysis of blinking fractals}
\label{s4}

We start by reminding   traditional definitions of the infinite
sequences and subsequences.  An \textit{infinite sequence}
$\{a_n\}, a_n \in A, n \in \mathbb{N},$ is a function having as
the domain the set of natural numbers, $\mathbb{N}$, and as the
codomain  a set $A$. A \textit{subsequence} is   a sequence from
which some of its elements have been cancelled. The IUA allows us
to prove the following result.
\begin{theorem}
\label{t2} The number of elements of any infinite sequence is less
or equal to~\ding{172}.
\end{theorem}

\textit{Proof.}  The IUA  states that the set $\mathbb{N}$ has
\ding{172} elements. Thus, due to the sequence definition given
above, any sequence having $\mathbb{N}$ as the domain  has
\ding{172} elements.

The notion of subsequence is introduced as a sequence from which
some of its elements have been cancelled. Thus, this definition
gives infinite sequences having the number of members less than
grossone.  \hfill $\Box$

It becomes appropriate now to define the \textit{complete
sequence} as an infinite sequence  containing \ding{172} elements.
For example, the sequence $\{n\}$ of natural numbers  is complete,
the sequences of even  and odd natural numbers  are not complete.

One of the immediate consequences of the understanding of this
result is that any sequential process can have at maximum
\ding{172} elements and (see \cite{Sergeyev,informatica}) it
depends on the chosen numeral system which numbers among
\ding{172} members of the process we can observe.

\begin{example}
\label{e12} Let us consider the set, $\widehat{\mathbb{N}}$, the
set of \textit{extended natural numbers} indicated as
$\widehat{\mathbb{N}}$ and including $\mathbb{N}$ as a proper
subset
 \beq
  \widehat{\mathbb{N}} = \{
1,2, \ldots ,\mbox{\ding{172}}-1, \mbox{\ding{172}},
\mbox{\ding{172}}+1, \ldots , \mbox{\ding{172}}^2-1,
\mbox{\ding{172}}^2, \mbox{\ding{172}}^2+1, \ldots \}.
\label{4.2.2}
       \eeq
        Then, starting from
the number 1, the process of the sequential counting can arrive at
maximum to \ding{172}
\[
\underbrace{1,2,3,4,\hspace{1mm}  \ldots \hspace{1mm}
\mbox{\ding{172}}-2,\hspace{1mm}
 \mbox{\ding{172}}-1,
\mbox{\ding{172}}}_{\mbox{\ding{172}}},  \mbox{\ding{172}}+1,
\mbox{\ding{172}}+2, \mbox{\ding{172}}+3, \ldots
\]
 Starting from 2 it    arrives at maximum
to $\mbox{\ding{172}}+1$
\[
1,\underbrace{2,3,4,\hspace{1mm}  \ldots \hspace{1mm}
\mbox{\ding{172}}-2,\hspace{1mm}
 \mbox{\ding{172}}-1,
\mbox{\ding{172}},  \mbox{\ding{172}}+1}_{\mbox{\ding{172}}},
\mbox{\ding{172}}+2, \mbox{\ding{172}}+3, \ldots
\]
 Starting from 3 it   arrives at maximum
to $\mbox{\ding{172}}+2$
\[
\begin{tabular}{cr}\hspace {20mm}$1,2,\underbrace{3,4,\hspace{1mm}  \ldots \hspace{1mm}
\mbox{\ding{172}}-2,\hspace{1mm}
 \mbox{\ding{172}}-1,
\mbox{\ding{172}},  \mbox{\ding{172}}+1,
\mbox{\ding{172}}+2}_{\mbox{\ding{172}}}, \mbox{\ding{172}}+3,
  \ldots$ &
\hspace {13mm}
 $\Box$
\end{tabular}
\]
 \end{example}

Similarly to infinite sets, the IUA imposes a more precise
description of infinite sequences. To define a sequence $\{a_n\}$
it is not sufficient just to give a formula for~$a_n$. It is
necessary   to indicate explicitly its number of elements.

\begin{example}
\label{e13} Let us consider the following three sequences,
$\{a_n\},\{b_n\},$ and $\{c_n\}$:
\[
 \{a_n\} =  \{ 2,\hspace{3mm} 4,\hspace{3mm} \ldots \hspace{3mm} 2(\mbox{\ding{172}}-1),\hspace{3mm}
2\mbox{\ding{172}} \},
\]
\beq \{b_n\}  = \{  2,\hspace{3mm}4,\hspace{3mm} \ldots
\hspace{3mm} 2 (\frac{2\mbox{\ding{172}}}{5}-1),\hspace{3mm}
2\cdot \frac{2\mbox{\ding{172}}}{5} \},
 \label{3.7.1}
       \eeq
\beq
 \{c_n\}  = \{ 2,\hspace{3mm} 4,\hspace{3mm} \ldots \hspace{3mm}
2 (\frac{4\mbox{\ding{172}}}{5}-1),\hspace{3mm} 2\cdot
\frac{4\mbox{\ding{172}}}{5} \}.
 \label{3.7.2}
       \eeq
They  have the same general element  equal to $2n$ but they  are
different because they have different number of members.  The
first sequence has \ding{172} elements and is thus complete,  the
other two sequences are not complete: $\{b_n\}$ has
$\frac{2\mbox{\ding{172}}}{5}$ elements  and $\{c_n\}$ has
$\frac{4\mbox{\ding{172}}}{5}$ members.  Note also that among
these three sequences only $\{b_n\}$ is a subsequence of the
sequence of even natural numbers because its last element has the
number $\frac{2\mbox{\ding{172}}}{5}  \le
\frac{\mbox{\ding{172}}}{2}$. Since \ding{172} is the last even
natural number, elements of $\{a_n\}$ and $\{c_n\}$ having $n >
\frac{\mbox{\ding{172}}}{2}$ are not natural but extended natural
numbers (see (\ref{4.2.2})). \hfill
 $\Box$
 \end{example}

Thus, to describe a sequence we should use the record $\{a_n: k
\}$ where $a_n$ is, as usual, the general element and $k$ is the
number (finite or infinite) of members of the sequence. In
connection to this definition the following natural question
arises inevitably. Suppose that we have two sequences, for
example, $\{b_n:\frac{2\mbox{\ding{172}}}{5}\}$ and
$\{c_n:\frac{4\mbox{\ding{172}}}{5}\}$ from (\ref{3.7.1}) and
(\ref{3.7.2}). Can we create a new sequence, $\{d_n:k\}$, composed
from both of them, for instance, as it is shown below
 \[
b_1,\hspace{1mm} b_2,\hspace{1mm} \ldots \hspace{1mm}
b_{\frac{2\mbox{\tiny{\ding{172}}}}{5}-2},\hspace{1mm}
b_{\frac{2\mbox{\tiny{\ding{172}}}}{5}-1},\hspace{1mm}b_{\frac{2\mbox{\tiny{\ding{172}}}}{5}},\hspace{1mm}
c_1,\hspace{1mm} c_2,\hspace{1mm} \ldots \hspace{1mm}
c_{\frac{4\mbox{\tiny{\ding{172}}}}{5}-2},\hspace{1mm}
c_{\frac{4\mbox{\tiny{\ding{172}}}}{5}-1},\hspace{1mm}
c_{\frac{4\mbox{\tiny{\ding{172}}}}{5}}
 \]
and which will be the value of the number of its elements $k$?

The answer is `no' because  due to the  definition of the infinite
sequence, a sequence can be at maximum complete,  i.e., it cannot
have more than $\mbox{\ding{172}}$  elements. Starting from the
element $b_1$ we can arrive at maximum to the element
$c_{\frac{3\mbox{\tiny{\ding{172}}}}{5}}$  being the element
number \ding{172} in the sequence $\{d_n:k\}$ which we try to
construct. Therefore, $k=\mbox{\ding{172}}$ and
\[
\underbrace{b_1,\hspace{1mm}  \ldots \hspace{1mm}
b_{\frac{2\mbox{\tiny{\ding{172}}}}{5}},\hspace{1mm}
 c_1,\hspace{1mm}  \ldots
c_{\frac{3\mbox{\tiny{\ding{172}}}}{5}}}_{\mbox{\ding{172}
elements}}, \hspace{1mm}
\underbrace{c_{\frac{3\mbox{\tiny{\ding{172}}}}{5}+1}, \ldots
\hspace{1mm}
c_{\frac{4\mbox{\tiny{\ding{172}}}}{5}}}_{\frac{\mbox{\tiny{\ding{172}}}}{5}
 \mbox{ elements }}.
\]
The remaining members of the sequence
$\{c_n:\frac{4\mbox{\ding{172}}}{5}\}$ will form the second
sequence, $\{g_n: l \}$ having $l=
\frac{4\mbox{\ding{172}}}{5}-\frac{3\mbox{\ding{172}}}{5} =
\frac{\mbox{\ding{172}}}{5}$  elements. Thus, we have formed two
sequences, the first of them is complete and the second is not.

The introduced more precise description of sequences allows us to
observe fractal processes at different points of infinity by
indicating the number of a step, $n, 1 \le n \le
\mbox{\ding{172}},$ that we want to study. For example, for our
blinking fractal from Fig.~\ref{Biology_3} we are able to say that
we observe grey squares at all odd steps and white circles at even
steps independently of the fact  $n$ is finite or infinite.
Particularly, since due to the Infinite Unit Axiom
$\frac{\mbox{\ding{172}}}{2}$ is integer, for $n=
\mbox{\ding{172}}$ we have white circles  and for $n=
\mbox{\ding{172}}-1$ -- grey squares.

In order to be able to measure fractals at infinity (e.g., to
calculate the number of squares or circles at a step $n$ in our
blinking  fractal from Fig.~\ref{Biology_3}), we should reconsider
the theory of divergent series from the new viewpoint introduced
in the previous sections. The introduced numeral system allows us
to express not only different finite numbers but also different
infinite numbers. Therefore    (see \cite{Sergeyev,informatica})
we should explicitly indicate the number of items in all sums
independently on the fact whether this number is finite or
infinite. We shall be able to calculate the sum  if its items, the
number of items, and the result are expressible in the numeral
system used for calculations. It is important to notice that even
though a sequence cannot have more than \ding{172} elements, the
number of items in a sum can be greater than grossone because the
process of summing up  not necessary should be executed by a
sequential addition of items.

For instance, let us consider two infinite series
$$S_1=1+2+4+8+16+\ldots, \hspace{15mm}S_2=1+2+1+2+1+2+1\ldots$$  The traditional
analysis gives us a very poor answer that both of them diverge to
infinity. Such operations as $S_1 - S_2$ or $ \frac{S_1}{S_2} $
are not defined. From the new point of view, the sums $S_1$ and
$S_2$ can be calculated  because it is necessary to indicate
explicitly the number of items in both   sums.

Suppose that the   sum $S_1$ has $m+1$ items and the sum $S_2$ has
$n$ items:
  \beq
S_1(m)=\underbrace{1+2+4+8+\ldots+2^m}_{m+1}, \hspace{6mm}
S_2(n)=\underbrace{1+2+1+2+1+\ldots}_n.
 \label{biology_1}
 \eeq
 Let us first calculate
the  sum  $S_1(m)$. It is evident that it is a particular case of
the sum
 \beq
 Q_m = \sum_{i=0}^{m}
q^i = 1 + q + q^2 + \ldots + q^m,
 \label{3.7.2.f}
 \eeq
where $m$ can be finite or infinite.      Traditional analysis
proves that  the geometric series $\sum_{i=0}^{\infty} q^i$
converges to $\frac{1}{1-q}$ for $q$ such that $-1 < q < 1$. We
are able to give a more precise answer for \textit{all} values of
$q$ and finite and infinite values of $m$.

By multiplying the left hand and the right hand parts of this
equality by $q$ and by subtracting the result from (\ref{3.7.2.f})
we obtain
\[
Q_m - qQ_m = 1-q^{m+1}
\]
and, as a consequence, for all $q\neq 1$ the formula
 \beq
  Q_m =
\frac{1-q^{m+1}}{1-q}
 \label{3.7.2.f.1}
 \eeq
holds for finite and infinite $m$. Thus, the possibility to
express infinite and infinitesimal numbers allows us   to take
into account    infinite $m$ too and the value $q^{m+1}$ being
infinitesimal for a finite $q<1$ and infinite for $q>1$. Moreover,
we can calculate $Q_m$ for $q=1$. In fact, in this case we have
just
\[
Q_m = \underbrace{1+1+1+\ldots+1}_{m+1} = m+1.
\]
Now, to calculate the sum   $S_1(m)$ it is sufficient to take
$q=2$
  \[
   S_1(m) = \underbrace{1+2+4+8+16+\ldots+2^m}_{m+1} =  \frac{1-2^{m+1}}{1-2}=2^{m+1}-1.
 \]
This formula can be used for finite and infinite values of $m$.
For instance, if $m=\frac{\mbox{\ding{172}}}{2}-1$ then
$S_1(\frac{\mbox{\ding{172}}}{2}-1)=2^{\frac{\mbox{\tiny{\ding{172}}}}{2}}-1$;
if $m=\frac{\mbox{\ding{172}}}{2}$ then
$S_1(\frac{\mbox{\ding{172}}}{2})=2^{\frac{\mbox{\tiny{\ding{172}}}}{2}+1}-1$.
Note that the sum $S_1(\frac{\mbox{\ding{172}}}{2})$ has been
obtained by adding $2^{\frac{\mbox{\tiny{\ding{172}}}}{2}}$ to the
sum $S_1(\frac{\mbox{\ding{172}}}{2}-1)$. In fact, if we subtract
from the obtained number
$2^{\frac{\mbox{\tiny{\ding{172}}}}{2}+1}-1$ this value, we obtain
exactly $S_1(\frac{\mbox{\ding{172}}}{2}-1)$:
\[
S_1(\frac{\mbox{\ding{172}}}{2}) -
2^{\frac{\mbox{\tiny{\ding{172}}}}{2}} =
2^{\frac{\mbox{\tiny{\ding{172}}}}{2}+1}-1 -
2^{\frac{\mbox{\tiny{\ding{172}}}}{2}} =
2^{\frac{\mbox{\tiny{\ding{172}}}}{2}} -1 =
S_1(\frac{\mbox{\ding{172}}}{2}-1).
\]

 The second sum, $S_2(n)$, from (\ref{biology_1}) is calculated as
 follows
\[
S_2(n)=\underbrace{1+2+1+2+1+\ldots}_{n} = \left \{
\begin{array}{ll} k+2k=3k, &
  \mbox{if  } n=2k,\\
k+2k+1=3k+1, &    \mbox{if  } n=2k+1.\\
 \end{array} \right.
\]
By giving numerical values (finite or infinite) to   $n$ we obtain
numerical values for results of the sum.  If, for instance,
$n=3\mbox{\ding{172}}$ then we obtain
$S_2(3\mbox{\ding{172}})=4.5\mbox{\ding{172}}$ because \ding{172}
is even. If     $n=3\mbox{\ding{172}}+1$ then  we obtain
$S_2(3\mbox{\ding{172}}+1)=4.5\mbox{\ding{172}}+1$. Note, that we
have no indeterminate expressions and the results such as $S_2(m)
- S_2(n),$ $S_1(m) - S_2(n),  $   $ \frac{S_2(m)}{S_2(n)},$ etc.
can be easily calculated.

Let us now return to fractals. First of all, it is evident that
the number of circles or squares at a step $i,  1 \le i \le
 \mbox{\ding{172}}$, in the blinking process from
Fig.~\ref{Biology_3}  is defined by the sum $S_1(i-1)$. It is
important to remind that, due to the IUA, a process cannot have
more than grossone steps but a sum can have more than grossone
items because it can be calculated in parallel (it is important
that it is not calculated in sequence). Thus, if we consider the
process from Fig.~\ref{Biology_3}, then in $S_1(i)$ it follows $ 1
\le i \le \mbox{\ding{172}}$.  It is possible to calculate
$S_1(i), i > \mbox{\ding{172}},$ if this is done without any
connection  to processes (i.e., $S_1(i)$ can be computed in
parallel) or in connection with another process with other
different initial conditions. For instance, starting the process
from Fig.~\ref{Biology_3} from two circles instead of one square,
it is possible to arrive to $S_1(\mbox{\ding{172}}+1),$ see
discussion in Example~\ref{e12}).

We conclude this section by calculating the side of the squares,
$s(i), i= 2k-1, 1 \le k \le \frac{\mbox{\ding{172}}}{2}$, and the
diameter of circles, $d(i), i= 2k, 1 \le k \le
\frac{\mbox{\ding{172}}}{2}$, for the blinking fractal from
Fig.~\ref{Biology_3}. It is easily to show that
\[
s(i)=\frac{1}{2^{k-1}},  \hspace{1cm} i= 2k-1,  \hspace{5mm} 1 \le
k \le \frac{\mbox{\ding{172}}}{2},
\]
\[
 d(i) = \frac{1}{2^{k}},  \hspace{1cm}
i= 2k, \hspace{5mm} 1 \le k \le \frac{\mbox{\ding{172}}}{2}.
\]
For  finite values of $i$ we obtain finite values of $s(i)$ and
$d(i)$, whereas for infinite values of $i$ we obtain infinitesimal
values of $s(i)$ and $d(i)$. For example, for
$i=\frac{\mbox{\ding{172}}}{3}$ it follows that we have white
circles (because due to the IUA, for all finite integer $n$ the
numbers of the form $\frac{\mbox{\ding{172}}}{n}$ are integer and,
therefore, $i=\frac{\mbox{\ding{172}}}{3}$ is even) and their
diameter $d(\frac{\mbox{\ding{172}}}{3})=
2^{-\frac{\mbox{\tiny{\ding{172}}}}{6}}$. Analogously, for
$i=\frac{\mbox{\ding{172}}}{3}-1$   we have grey squares having
the side $s(\frac{\mbox{\ding{172}}}{3}-1)=
 2^{1-\frac{\mbox{\tiny{\ding{172}}}}{6}}$.

 Thus, the new infinite and infinitesimal numerals allow us to
 observe and to measure the traditional and blinking fractals at different points
 of infinity.

\section{A blinking fractals model of the growth of a forest and its analysis at infinity  }
\label{s5}

\begin{figure}[t]
   \vspace*{-7mm}\begin{center}
      \epsfig{ figure = 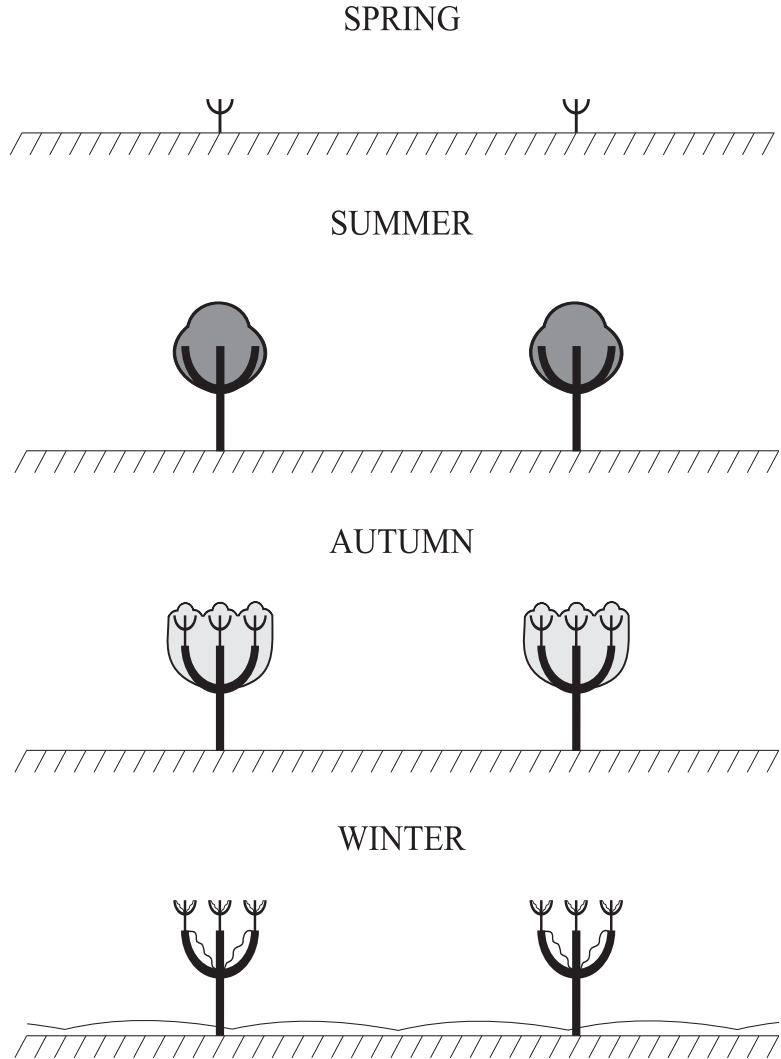, width = 4in, height = 5.5in,  silent = yes }
    \caption{The first year of growth.}
 \label{Biology_6}
  \end{center}
\end{figure}

The analysis performed in the previous section allows us to pass
to modeling processes in nature having a blinking fractals
structure. Without loss of the generality we consider biennial
plants (called hereinafter for simplicity \textit{trees}) that
will be observed four times a year: in spring, in  summer, in
autumn, and in winter. The process of growth starts in spring by
planting two small trees (see Fig.~\ref{Biology_6}, spring) in a
line. In summer, the trees grow up and green leafs (shown by grey
color) appear. During summer new branches appear and when we
observe our trees in autumn, we see these new branches and leafs
that meanwhile have became yellow (shown in Fig.~\ref{Biology_6}
by the light grey color). When we observe our small forest in
winter, there are no leafs and the trees are under the snow.

\begin{figure}[t]
  \vspace*{-7mm}\begin{center}
      \epsfig{ figure = 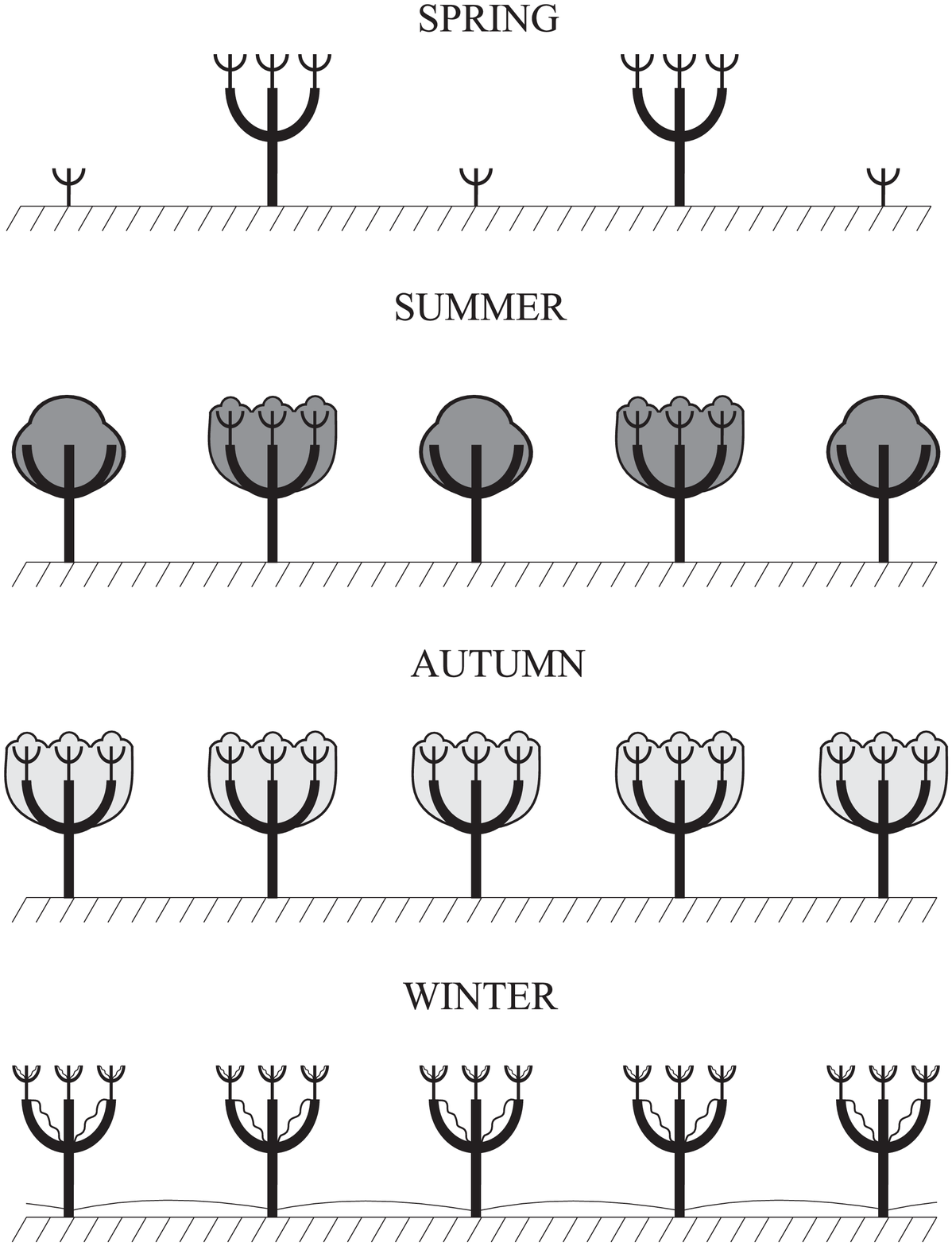, width = 4in, height = 5.5in,  silent = yes }
    \caption{The second year of growth.}
 \label{Biology_7}
  \end{center}
\end{figure}

\begin{figure}[ht]
  \begin{center}
      \epsfig{ figure = 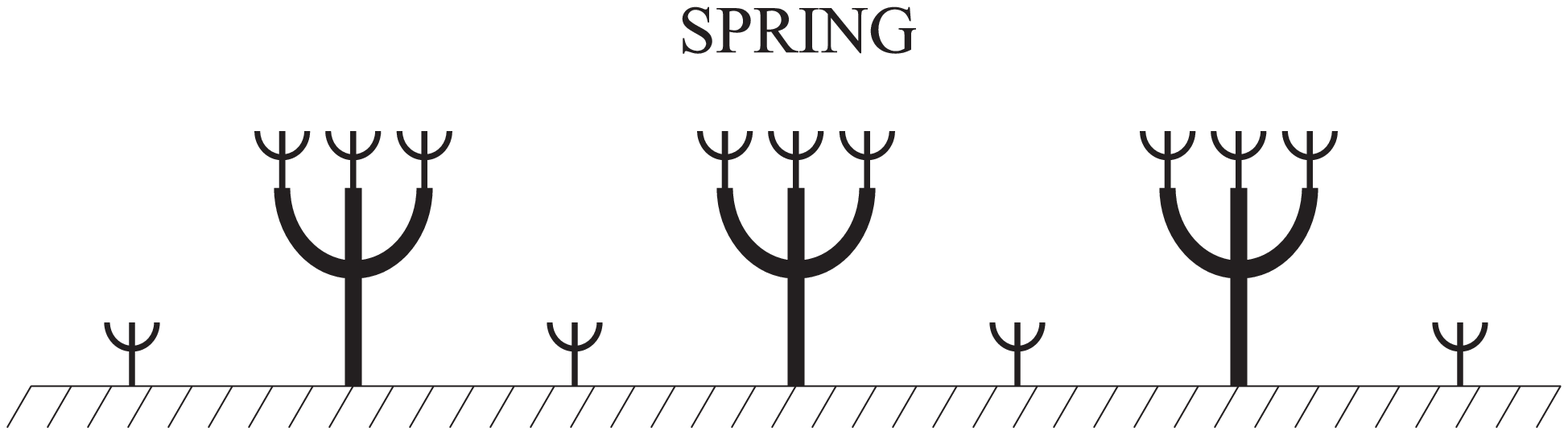, width = 4in, height = 1.1in,  silent = yes }
    \caption{Spring of the third year of growth.}
 \label{Biology_8}
  \end{center}
\end{figure}

When we   observe   our forest in spring of the second year (see
Fig.~\ref{Biology_7}, spring), we see that three new trees have
appeared (we suppose that the growth goes along the line defined
by the first two trees). In summer of the second year, we see
green leafs and observe that the new trees have grown up but the
old two trees are not able to grow up and remain the same. In
autumn, all five trees have the same measure and yellow leafs. In
winter, all of them are under the snow. During the winter two old
trees die and at their places new young trees appear. Two more new
trees appear also at the free places on the left and the right
ends of our forest. Thus, in spring of the third year we observe
the situation shown in Fig.~\ref{Biology_8}.

\begin{figure}[ht]
  \vspace*{-2mm}\begin{center}
      \epsfig{ figure = 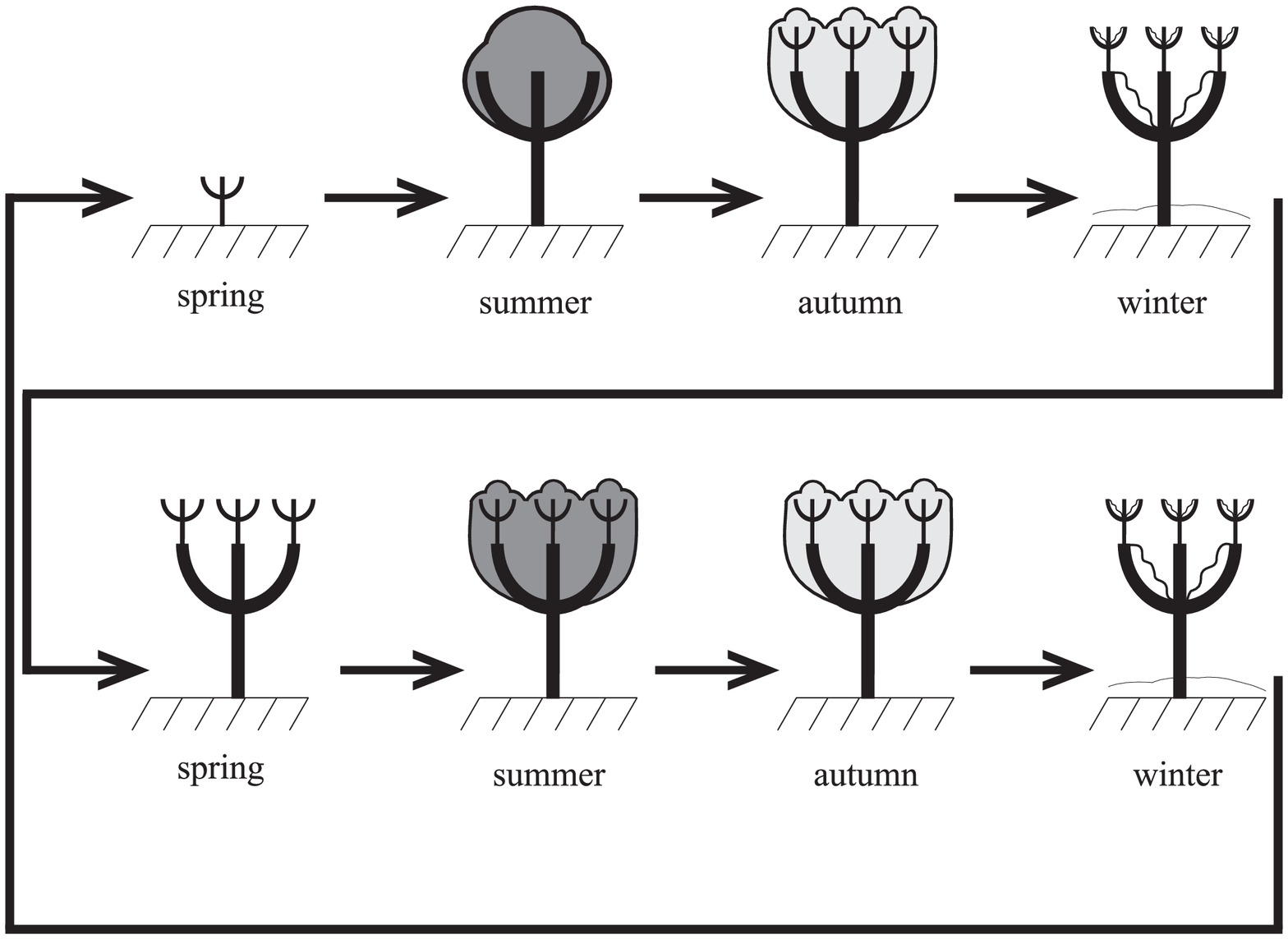, width = 5in, height = 3.4in,  silent = yes }
    \caption{The two years cycle of growth.}
 \label{Biology_9}
  \end{center}
\end{figure}

The process then continues to infinity and at each place where a
tree appears we can observe the two years cycle shown in
Fig.~\ref{Biology_9}. It is evident that the described model is
not a fractal. By using traditional mathematics we are not able to
answer to the following questions: How many trees and how many
branches will have our forest at infinity? What will be color of
the leafs? However, we can separate from the process of growth
several processes behaving as   fractals and, as a consequence,
the process of growth of our forest is a blinking fractal. The new
approach introduced in the previous section will allow us to give
quantitative answers to the questions above easily.

\begin{figure}[t]
  \begin{center}
      \epsfig{ figure = 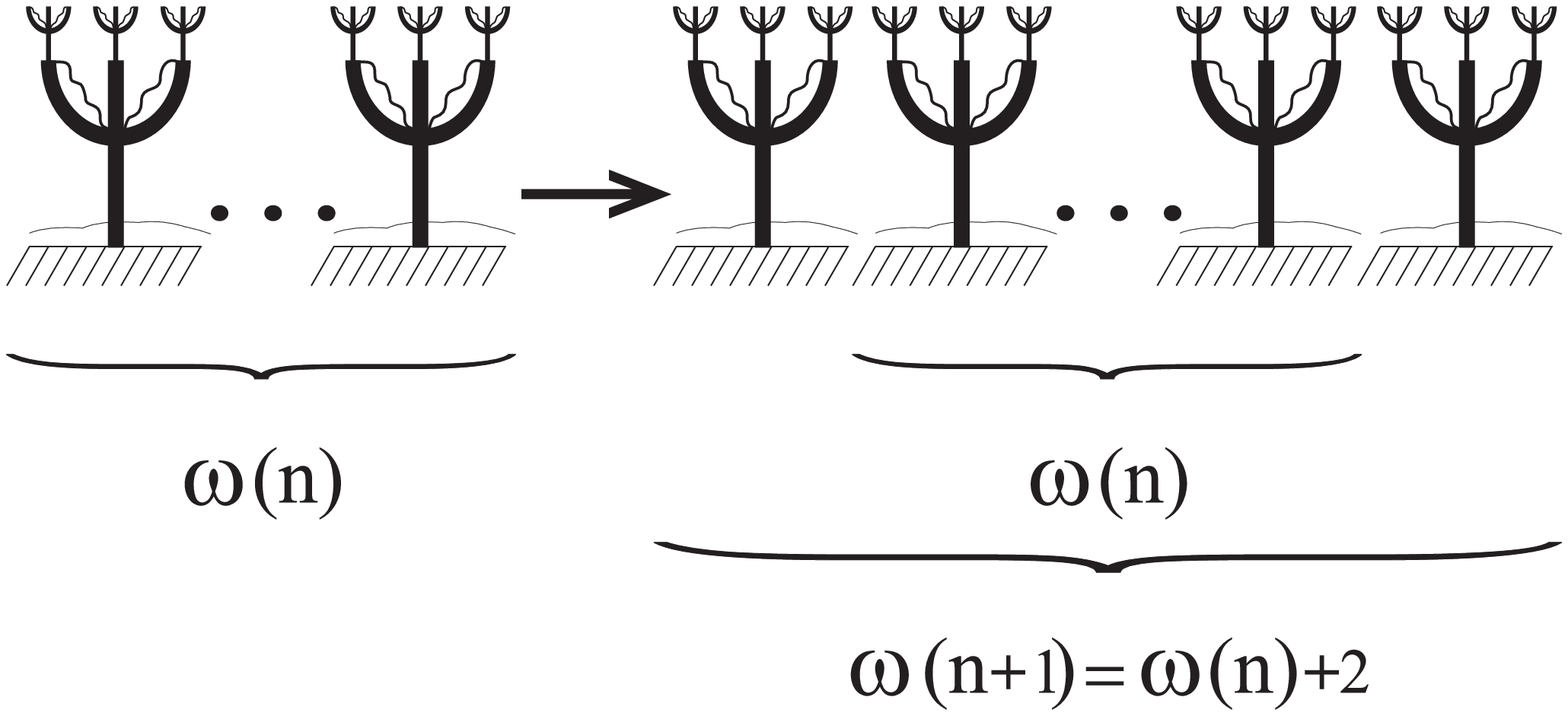, width = 3.5in, height = 1.6in,  silent = yes }
    \caption{The winter process.}
 \label{Biology_10}
  \end{center}
\end{figure}

\begin{figure}[t]
  \begin{center}
      \epsfig{ figure = 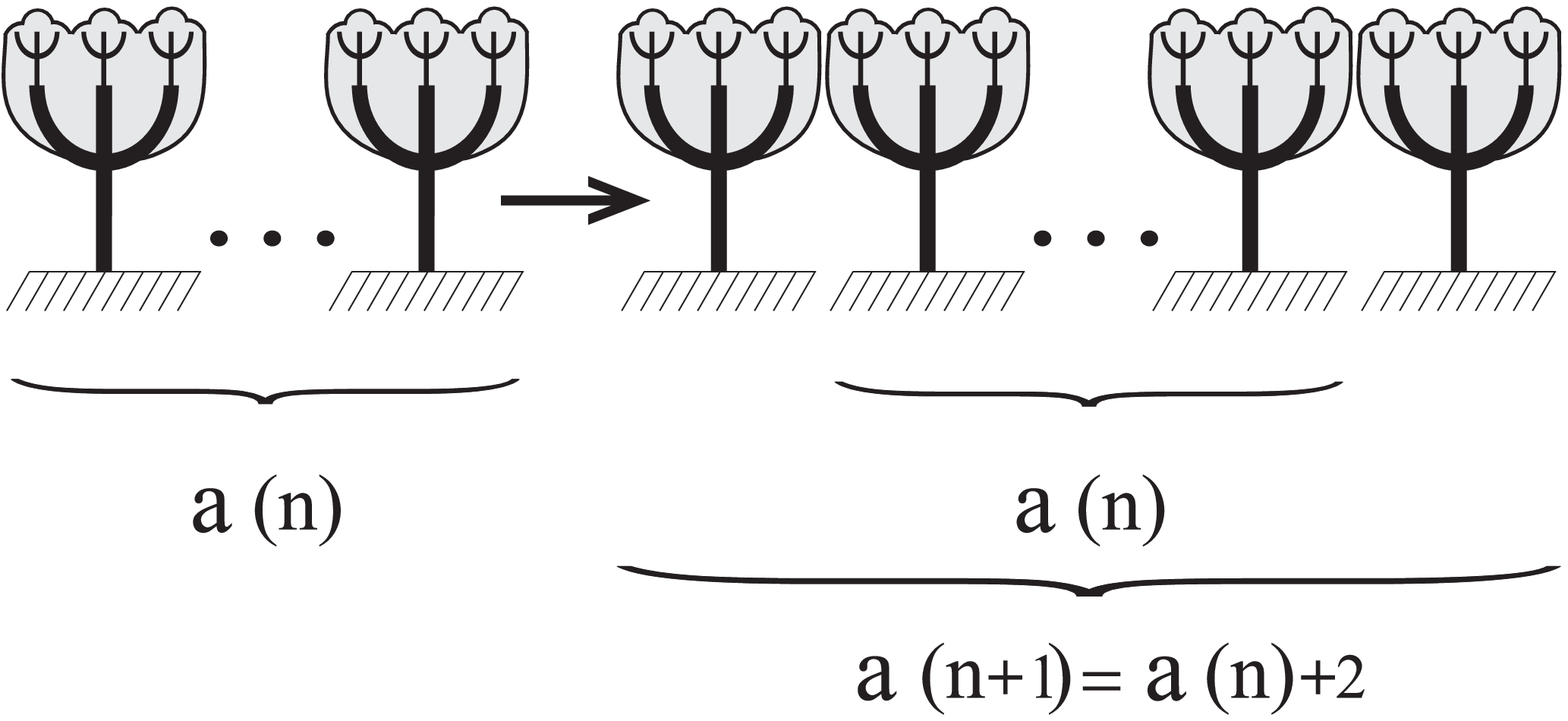, width = 3.5in, height = 1.6in,  silent = yes }
    \caption{The autumn process.}
 \label{Biology_11}
  \end{center}
\end{figure}

It is evident that during each season we have different basic
figures and it is necessary to consider the forest at each season
separately by applying the methodology of blinking fractals. Let
us start from winter. If $n$ is the number of the year then (see
Fig.~\ref{Biology_10}) we can easily calculate the number of trees
at the forest, $\omega(n)$,  for each year $n$ as follows
\[
\omega(1)=2, \,\,\,\, \omega(2)=5,\,\,\,\, \omega(3)=7, \ldots
\,\,\,\, \omega(n)=2n+1,  \,\,\,\,n \ge 2.
\]
An analogous formula for calculating the number of the trees in
autumn, $a(n)$, can be obtained (see Fig.~\ref{Biology_11}) for
the autumn process
\[
a(1)=2, \,\,\,\, a(2)=5, \,\,\,\, a(3)=7, \ldots \,\,\,\,
a(n)=2n+1, \,\,\,\, n \ge 2.
\]
The same formulae for calculating the number of trees   can be
obtained for spring and summer, because the number of trees is the
same during four seasons of each fixed year. Note that since we
observe our forest four time per year and any process, due to the
IUA, cannot have more than \ding{172} elements, the following
restriction exists for the number, $n$, of the years of
observations of our forest: $1 \le n \le
\frac{\mbox{\ding{172}}}{4}$. This means, particularly,  that the
number of the trees at the last year
$a(\frac{\mbox{\ding{172}}}{4})= \frac{\mbox{\ding{172}}}{2}+1$.

\begin{figure}[t]
  \begin{center}
      \epsfig{ figure = 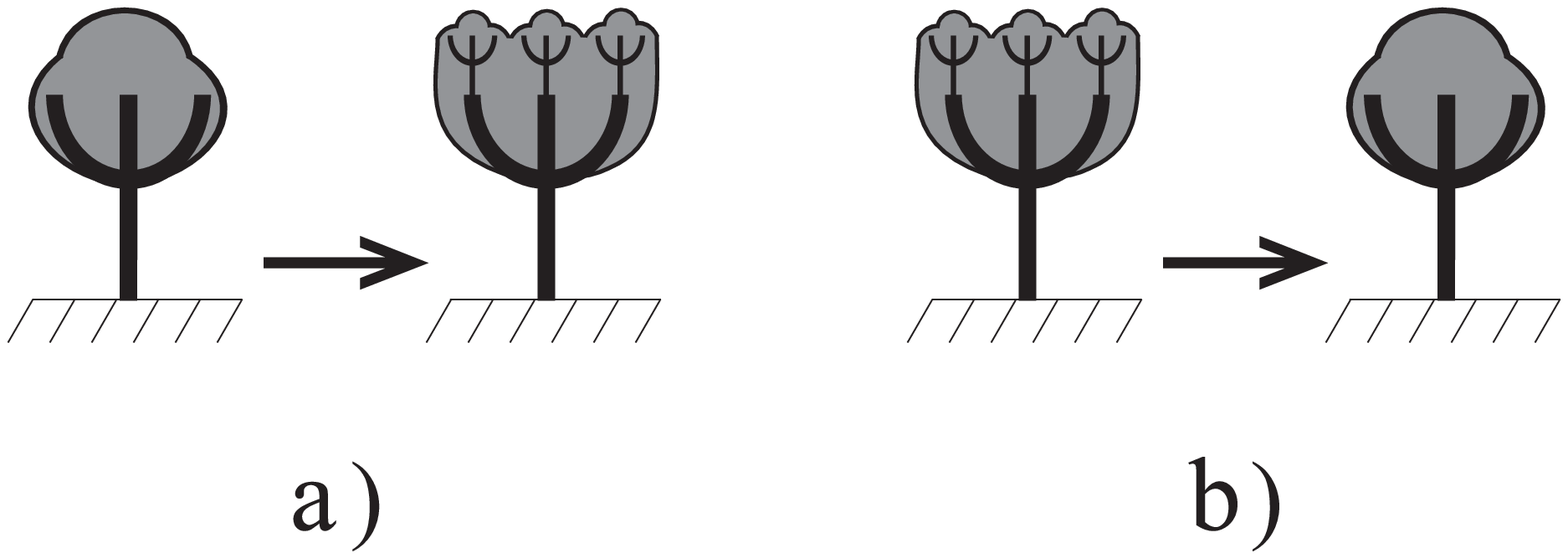, width = 3.4in, height = 1.2in,  silent = yes }
    \caption{Two summer processes.}
 \label{Biology_12}
  \end{center}
\end{figure}

\begin{figure}[t]
  \begin{center}
      \epsfig{ figure = 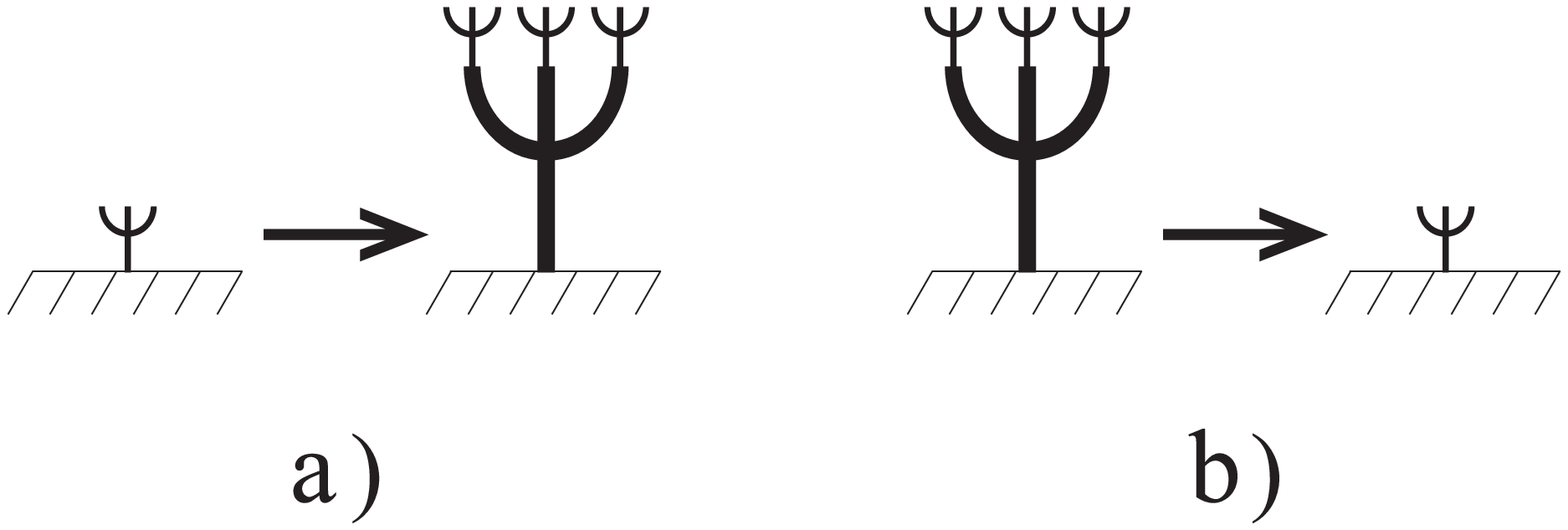, width = 3.4in, height = 1.1in,  silent = yes }
    \caption{Two spring processes.}
 \label{Biology_13}
  \end{center}
\end{figure}

Although the number of the trees does not change during each fixed
year, the changes take place for the number of branches;   the
form of the basic figures is also different for each season.
Moreover, as it emphasized in Figs.~\ref{Biology_12}
and~\ref{Biology_13}, the processes in summer and spring are more
complex than the processes in winter and autumn because we can
distinguish the processes of growth of the same tree (parts
\textit{a)} in both figures) and the process of substitution of an
old tree by the young one (shown in parts \textit{b)} in both
figures). For these two seasons we are able to calculate the
number of old trees, the number of young trees, and even the
number of branches in the forest for each of four seasons.

Let us first calculate the number of young and old  trees in
summer indicated as $y(i)$  and $o(i)$, respectively, that there
are in the forest during an observation $i$ made in the year $n$
(as it can be seen from Figs.~\ref{Biology_6} and~\ref{Biology_7},
in spring we have the same quantities that can be calculated by a
complete analogy). Since it can be done at maximum grossone
observations, it follows from description of the process of the
growth that for years $n=1,2,3,\,\, \ldots ,
\frac{\mbox{\ding{172}}}{4},$   the numbers of observations
corresponding to summer are $i=2,6,10, \,\, \ldots, \,\, 2+4(n-1),
\ldots, \mbox{\ding{172}}-2$. Thus, the numbers of young and old
trees in summer are calculated as follows
 \beq
  y(2+4(n-1)) = n+1, \hspace{10mm} 1 \le n
\le \frac{\mbox{\ding{172}}}{4}, \label{biology_2}
       \eeq
 \beq
 o(2) = 0, \hspace{5mm}   o(2+4(n-1)) = n, \hspace{10mm} 2 \le n
\le \frac{\mbox{\ding{172}}}{4}.
 \label{biology_3}
       \eeq

Let us calculate now the number of the branches in the forest for
each season. We indicate this number as $b(i)$ where $i$ is the
number of observation.  In winter, all the trees have the same
number of branches: three big and nine small. The observations
$i=4n$ correspond to winter at the year $n$ and there are $2n+1$
trees in the forest. Thus, we obtain $b(4n)=12(2n+1)$.
Analogously, the observations $i=3+4(n-1)$ correspond to autumn at
the year $n$ and, consequently, the number of branches in autumn
$b(3+4(n-1))=12(2n+1)$ as well.

In spring and summer, the situation is different: young and old
trees have different number of branches (see
Figs.~\ref{Biology_6},\ref{Biology_7},\ref{Biology_12},
and~\ref{Biology_13}). Let us consider summer (spring can be
studied by a complete analogy). In summer young trees have three
big branches each and old trees have 12 branches each: three big
and nine small.

Remind, that the observations  at the year $n$ corresponding to
summer have numbers $i=2+4(n-1)$. Thus, in order to obtain the
required result it is sufficient to use formulae (\ref{biology_2})
and (\ref{biology_3})  that give us
\[
b(2)=6, \hspace{5mm} b(2+4(n-1))= 3(n+1)+9n =12n+3, \hspace{10mm}
2 \le n \le \frac{\mbox{\ding{172}}}{4}.
\]
For example, in summer of the last possible year of observation,
$n=\frac{\mbox{\ding{172}}}{4}$, our forest has $
3\mbox{\ding{172}}+3$ branches.

\section{Concluding remarks}
\label{s6}

Fractals have been widely used in literature to model complex
systems. In this paper, a new type of objects -- blinking fractals
-- that are not covered by traditional theories studying
self-similarity processes have been used for studying season
changes during the growth of biological systems. The behavior of
blinking fractals at infinity has been investigated using infinite
and infinitesimal numbers proposed recently and a number of
quantitative characteristics    has been obtained.

As an example of application of the developed mathematical tools
for describing the behavior of complex biological systems a new
model of growth of a forest has been introduced and investigated
using the notion of the blinking fractals. The new mathematical
tools introduced in the paper have allowed us to separate in this
complex model several fractal processes and to perform their
accurate quantitative analysis. It is evident that the introduced
model can be easily generalized to describe more complex objects
and systems. For example, it is possible to introduce plants with
the cycle  of life superior to two years, the plants having a more
complex structure can be also described by the introduced
approach. In the future it is possible also to study some
additional mechanisms (for instance, plant pests or nature
disasters) that influence the process of growth.

\bibliographystyle{plain}
\bibliography{XBib_Biology}
\end{document}